\documentclass[lettersize,journal]{IEEEtran}
\usepackage{amsmath,amsfonts}
\usepackage{amssymb}
\usepackage{algorithmic}
\usepackage{algorithm}
\usepackage{array}
\usepackage{tikz}
\usepackage{pgfplots}
\usepackage[caption=false,font=normalsize,labelfont=sf,textfont=sf]{subfig}
\usepackage{textcomp}
\usepackage{stfloats}
\usepackage{url}
\usepackage{soul}
\usepackage{verbatim}
\usepackage{graphicx}
\usepackage{cite}
\usepackage{caption}
\begin{document}
\thispagestyle{empty}
\pagestyle{empty}
\tikzset{new spy style/.style={spy scope={%
				magnification=2.8,
				size=1.25cm,
				connect spies,
				every spy on node/.style={
					rectangle,
					draw,
				},
				every spy in node/.style={
					draw,
					rectangle,
				}
			}
		}
	}

\title{Tertiary-Mode STAR-RIS for Secure NOMA: Integrating Transmission, Reflection, and Jamming}

\author{
\small 
\IEEEauthorblockN{Mansi Nema\IEEEauthorrefmark{1}, Kuntal Deka\IEEEauthorrefmark{1},  Sanjeev Sharma\IEEEauthorrefmark{2}, and Tharmalingam Ratnarajah\IEEEauthorrefmark{2}}\\
		\IEEEauthorblockA{\IEEEauthorrefmark{1}Department of Electronics \& Electrical Engineering, Indian Institute of Technology Guwahati, India\\\IEEEauthorrefmark{2}Department of Electronics Engineering, Indian Institute of Technology (BHU) Varanasi, India\\ \IEEEauthorrefmark{3}Dept. of Electrical and Computer Engg., San Diego State University, San Diego,  USA\\
			emails: \{n.mansi, kuntaldeka\}@iitg.ac.in,  sanjeev.ece@iitbhu.ac.in, t.ratnarajah@ieee.org 
            \vspace{-0.4in}
	}}



\maketitle
\thispagestyle{empty}
\pagestyle{empty}
\begin{abstract}
This paper investigates the physical layer security of a non-orthogonal multiple access (NOMA) system assisted by a tertiary-mode simultaneously transmitting and reflecting reconfigurable intelligent surface (STAR-RIS), which can perform transmission, reflection, and jamming simultaneously. The  system comprises a base station (BS) serving two users located on opposite sides of the STAR-RIS, assuming perfect channel state information (CSI) at the transmitter. To enhance secrecy performance, a subset of STAR-RIS elements is adaptively configured for jamming. A penalty-based alternating optimization algorithm is developed to jointly optimize the BS’s active beamforming and the STAR-RIS’s passive beamforming and mode selection. Simulation results demonstrate that the proposed design substantially improves the achievable sum rate and secrecy performance compared to conventional RIS-assisted and no-RIS benchmarks, highlighting the potential of tertiary-mode STAR-RIS for secure and efficient next-generation wireless communications.

\end{abstract}

\begin{IEEEkeywords}
Physical layer security, NOMA, RIS.
\end{IEEEkeywords}

\vspace{-0.1in}
\section{Introduction}
\IEEEPARstart{R}econfigurable intelligent surfaces (RIS) have emerged as a promising technology for next-generation wireless communication systems. They can significantly enhance both spectral and energy efficiency. An RIS consists of many low-cost passive elements that can adjust the amplitude and phase of incident electromagnetic (EM) waves. By properly configuring these elements, the RIS can reconfigure the propagation environment to strengthen desired signals and suppress interference \cite{5}. Unlike active relays, RIS does not require power amplifiers or radio-frequency (RF) chains, which makes it simple, cost-effective, and energy-efficient. Owing to these advantages, RIS is recognized as an important enabler for sustainable and energy-efficient sixth-generation (6G) wireless networks.

Although RIS offers many benefits, traditional implementations operate only in the reflection mode and can serve users located on a single side of the surface. This directional limitation restricts its capability in providing full-space coverage, particularly in dense urban or non-line-of-sight (NLoS) environments. To overcome this limitation, the concept of a simultaneously transmitting and reflecting RIS (STAR-RIS) has been introduced \cite{4}. A STAR-RIS can both transmit and reflect incident signals, extending its coverage to users on both sides of the surface. 

In recent years, STAR-RIS has also attracted growing attention in the context of physical layer security (PLS) and its integration with non-orthogonal multiple access (NOMA) \cite{1,3,6}. These studies demonstrate that STAR-RIS can simultaneously enhance coverage, improve spectral efficiency, and increase the secrecy capacity of wireless networks. However, achieving optimal performance requires efficient configuration of the STAR-RIS, which remains a major challenge. The associated optimization problems are highly non-convex and involve the joint design of beamforming, phase adjustment, and mode selection. To address these challenges, several works have proposed iterative optimization frameworks such as alternating optimization (AO) and penalty-based methods \cite{2}, \cite{7}, \cite{8},  \cite{12}. These approaches achieve a reasonable balance between system performance and computational complexity.

Despite these advances, most existing STAR-RIS studies employ energy splitting (ES) or time switching (TS) protocols, where the transmitting and reflecting elements are separated in the energy or time domain. Although ES and TS are simple to implement, they limit the flexibility and adaptability of the RIS elements, particularly in dynamic wireless environments.

This paper adopts a more versatile mode switching (MS) protocol, where each STAR-RIS element can independently operate in transmission, reflection, or jamming mode according to network and security requirements. We consider a STAR-RIS-assisted multiple-input single-output (MISO) NOMA system in which a multi-antenna base station (BS) communicates with two legitimate users located on opposite sides of the STAR-RIS in the presence of an eavesdropper. The ability of each STAR-RIS element to dynamically select its operating mode provides fine-grained control over the propagation environment, enhancing both throughput and secrecy. To the best of the authors’ knowledge, this is the first work to analyze the secrecy and sum-rate performance of a tertiary-mode STAR-RIS-assisted NOMA system.

The main contributions of this paper are:
\begin{itemize}
\item A secure STAR-RIS-assisted MISO-NOMA communication framework is proposed, where each STAR-RIS element can dynamically operate in transmission, reflection, or jamming mode to jointly enhance the achievable sum-rate and physical layer security.
\item A joint optimization problem is formulated to maximize the system sum-rate by jointly designing the active beamforming at the BS and the passive beamforming and mode selection at the STAR-RIS.

\item A penalty-based alternating optimization algorithm is developed to efficiently solve the resulting non-convex problem, which simultaneously handles continuous and discrete optimization variables in a unified framework.


\end{itemize}
\vspace{-0.1in}
\section{System Model and Problem Formulation}

\begin{figure}[!htb]
    \centering
\includegraphics[width=0.8 \linewidth]{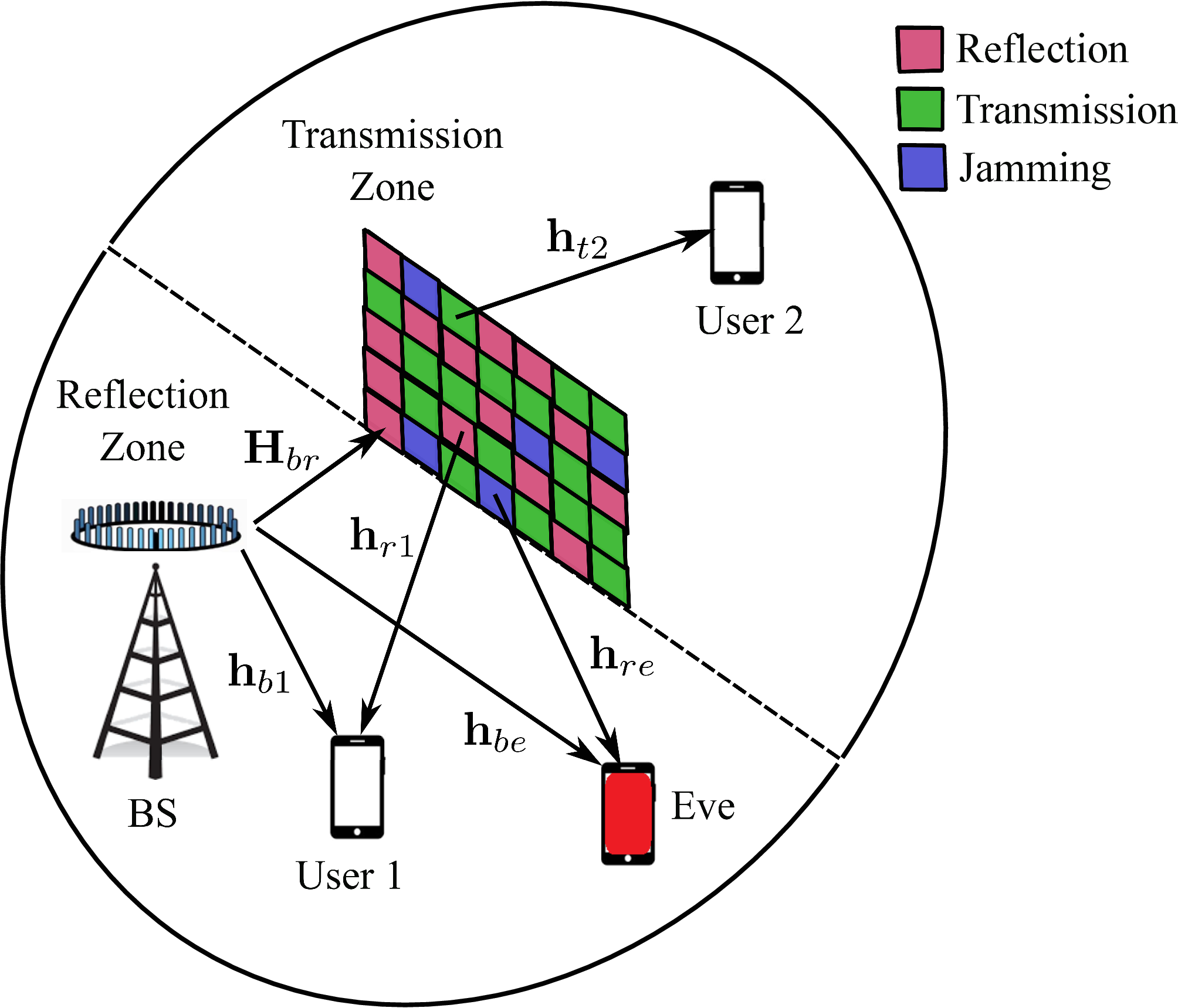}
\caption{STAR-RIS assisted MISO NOMA communication system.}
\label{fig_1}
\end{figure}

Consider a tertiary-mode MISO STAR-RIS system for secure NOMA communications, as illustrated in Fig.~\ref{fig_1}. The BS, equipped with 
$N$ transmit antennas, serves two legitimate users: one located on the transmission side and the other on the reflection side of the STAR-RIS. Each user is equipped with a single receive antenna. In addition, a passive eavesdropper is positioned on the reflection side of the STAR-RIS. It is assumed that the eavesdropper cannot intercept the signal intended for the transmission-side user because there is no direct link between them. Furthermore, the transmission-side user does not have a line-of-sight (LoS) channel with the BS.

\vspace{-0.1in}

\vspace{-0.15in}
\subsection{Channel Model}
Let \( \mathbf{H}_{\rm{br}} \in \mathbb{C}^{K \times N} \) denote the channel matrix from the BS to the STAR-RIS, where \( K \) represents the number of STAR-RIS elements. 
The channels from the STAR-RIS to User~1, User~2, and Eve are denoted by \( \mathbf{h}_{\rm{r1}} \in \mathbb{C}^{K \times 1} \), \( \mathbf{h}_{\rm{t2}} \in \mathbb{C}^{K \times 1} \), and \( \mathbf{h}_{\rm{re}} \in \mathbb{C}^{K \times 1} \), respectively. 
The direct links between the BS and User~1 and between the BS and Eve are represented by \( \mathbf{h}_{\rm{b1}} \in \mathbb{C}^{N \times 1} \) and \( \mathbf{h}_{\rm{be}} \in \mathbb{C}^{N \times 1} \), respectively. 
The direct BS–User~2 link is assumed to be negligible due to severe path loss or blockage, which is typical in STAR-RIS-assisted communication scenarios. User~1 is nearer to the BS as shown in Fig.~\ref{fig_1}.
\vspace{-0.2cm}
\subsection{Signal Model}
The transmitted signal from the BS is expressed as
$ \mathbf{x} = \mathbf{w}_1 s_1 + \mathbf{w}_2 s_2 $, where \( s_1 \) and \( s_2 \) denote the information-bearing symbols intended for User~1 and User~2, respectively, satisfying \( \mathbb{E}[|s_1|^2] = \mathbb{E}[|s_2|^2] = 1 \). 
Here, \( \mathbf{w}_1 \) and \( \mathbf{w}_2 \in \mathbb{C}^{N \times 1} \) represent the active beamforming vectors for the two users. 
The total transmit power at the BS is constrained by \( P_{\mathrm{max}} \), such that
$\|\mathbf{w}_1\|^2 + \|\mathbf{w}_2\|^2 \leq P_{\mathrm{max}} $.
The received signals at User~1, User~2, and Eve are:
\begin{align}
y_1 &= (\mathbf{h}_{\rm{r1}}^H \boldsymbol{\Theta}_{\rm r} \mathbf{H}_{\rm{br}} + \mathbf{h}_{\rm{b1}}^H)(\mathbf{w}_1 s_1 + \mathbf{w}_2 s_2) + \mathbf{h}_{\rm{r1}}^H \mathbf{z} + n_1, \\
y_2 &= \mathbf{h}_{\rm{t2}}^H \boldsymbol{\Theta}_{\rm t} \mathbf{H}_{\rm{br}} (\mathbf{w}_1 s_1 + \mathbf{w}_2 s_2) + n_2, \\
y_e &= (\mathbf{h}_{\rm{re}}^H \boldsymbol{\Theta}_{\rm r} \mathbf{H}_{\rm{br}} + \mathbf{h}_{\rm{be}}^H)(\mathbf{w}_1 s_1 + \mathbf{w}_2 s_2) + \mathbf{h}_{\rm{re}}^H \mathbf{z} + n_e,
\end{align}
where \( n_1, n_2, \text{and } n_e \sim \mathcal{CN}(0, \sigma^2) \) denote the independent additive white Gaussian noise (AWGN) at User~1, User~2, and Eve, respectively. 
The reflected jamming signal vector is denoted by \( \mathbf{z} = \boldsymbol{\Theta}_{\rm j} \mathbf{H}_{\rm{br}} \mathbf{x} \in \mathbb{C}^{K \times 1} \).

\vspace{-0.1in}
\subsection{STAR-RIS Operation Modes}
Each STAR-RIS element can operate in one of three mutually exclusive modes: reflection, transmission, or jamming. 
The corresponding diagonal phase-shift matrices are:
\begin{align}
\boldsymbol{\Theta}_{\rm r} &= \mathrm{diag}(\mathbf{r}), \quad 
\mathbf{r} = [\beta_{r,1} e^{j\theta_{r,1}}, \dots, \beta_{r,K} e^{j\theta_{r,K}}]^H, \\
\boldsymbol{\Theta}_{\rm t} &= \mathrm{diag}(\mathbf{t}), \quad 
\mathbf{t} = [\beta_{t,1} e^{j\theta_{t,1}}, \dots, \beta_{t,K} e^{j\theta_{t,K}}]^H, \\
\boldsymbol{\Theta}_{\rm j} &= \mathrm{diag}(\mathbf{j}), \quad 
\mathbf{j} = [\beta_{j,1} e^{j\theta_{j,1}}, \dots, \beta_{j,K} e^{j\theta_{j,K}}]^H,
\end{align}
where \( \beta_{r,k}, \beta_{t,k}, \beta_{j,k} \in \{0,1\} \) denote the mode-selection coefficients for the \( k \)-th element, indicating whether it operates in reflection, transmission, or jamming mode. 
To ensure exclusive operation, each element satisfies \( \beta_{r,k} + \beta_{t,k} + \beta_{j,k} = 1 \). 
The corresponding phase shifts \( \theta_{r,k} \), \( \theta_{t,k} \), and \( \theta_{j,k} \) are continuous variables within the range \( [0, 2\pi) \). We use this simplified model to keep the optimization problem tractable. Extending the analysis to a more general model with multiple active modes is an interesting direction for future work.

\vspace{-0.1in}
\subsection{SINR and Achievable Rates}
Using the above model, the various signal-to-interference-plus-noise ratio (SINR) term at each receiver can be expressed as follows.

For User~1 decoding its own signal:
\begin{equation}
\mathrm{SINR}_{11} = 
\frac{\left| (\mathbf{h}_{\rm{r1}}^H \boldsymbol{\Theta}_{\rm r} \mathbf{H}_{\rm{br}} + \mathbf{h}_{\rm{b1}}^H) \mathbf{w}_1 \right|^2}{
\left| (\mathbf{h}_{\rm{r1}}^H \boldsymbol{\Theta}_{\rm r} \mathbf{H}_{\rm{br}} + \mathbf{h}_{\rm{b1}}^H) \mathbf{w}_2 \right|^2 
+ \left| \mathbf{h}_{\rm{r1}}^H \boldsymbol{\Theta}_{\rm j} \mathbf{H}_{\rm{br}} \mathbf{w}_1 \right|^2 + \sigma^2}.
\label{eq:7}
\end{equation}

For User~2 decoding its own signal:
\begin{equation}
\mathrm{SINR}_{22} = 
\frac{\left| (\mathbf{h}_{\rm{t2}}^H \boldsymbol{\Theta}_{\rm t} \mathbf{H}_{\rm{br}}) \mathbf{w}_2 \right|^2}{
\left| (\mathbf{h}_{\rm{t2}}^H \boldsymbol{\Theta}_{\rm t} \mathbf{H}_{\rm{br}}) \mathbf{w}_1 \right|^2 + \sigma^2}.
\label{eq:71}
\end{equation}

For User~1 decoding User~2’s signal (SIC process):
\begin{equation}
\mathrm{SINR}_{12} = 
\frac{\left| (\mathbf{h}_{\rm{r1}}^H \boldsymbol{\Theta}_{\rm r} \mathbf{H}_{\rm{br}} + \mathbf{h}_{\rm{b1}}^H) \mathbf{w}_2 \right|^2}{
\left| (\mathbf{h}_{\rm{r1}}^H \boldsymbol{\Theta}_{\rm r} \mathbf{H}_{\rm{br}} + \mathbf{h}_{\rm{b1}}^H) \mathbf{w}_1 \right|^2 
+ \left| \mathbf{h}_{\rm{r1}}^H \boldsymbol{\Theta}_{\rm j} \mathbf{H}_{\rm{br}} \mathbf{w}_1 \right|^2 + \sigma^2}.
\label{eq:73}
\end{equation}

For Eve decoding User~1’s signal:
\begin{equation}
\mathrm{SINR}_{\rm{e1}} =
\frac{\left| (\mathbf{h}_{\rm{re}}^H \boldsymbol{\Theta}_{\rm r} \mathbf{H}_{\rm{br}} + \mathbf{h}_{\rm{be}}^H) \mathbf{w}_1 \right|^2}{
\left| \mathbf{h}_{\rm{re}}^H \boldsymbol{\Theta}_{\rm j} \mathbf{H}_{\rm{br}} \mathbf{w}_1 \right|^2 + \left| \mathbf{h}_{\rm{re}}^H \boldsymbol{\Theta}_{\rm j} \mathbf{H}_{\rm{br}} \mathbf{w}_2 \right|^2+\sigma^2}.
\label{eq:8}
\end{equation}

The corresponding achievable rates (in bits per second per hertz) for the legitimate user and Eve are given by
\begin{align}
R_{12} &= \log_2(1 + \mathrm{SINR}_{12}), \\
R_{\rm{e1}} &= \log_2(1 + \mathrm{SINR}_{\rm{e1}}).
\end{align}

\vspace{-0.2in}
\subsection{Problem Formulation}
Our objective is to jointly optimize the BS beamforming vectors and the STAR-RIS configuration to maximize the total achievable sum-rate of both users while maintaining secure transmission against Eve. Let $ \mathcal{K} $ denote the set of  the RIS element indices i.e. $ \mathcal{K} = \{1,2,\dots, {K}\} $. The optimization problem can thus be formulated as:
\begin{subequations}\label{eq:opt_problem}
\begin{align}
(\mathrm{P1}):\;
\max_{\mathbf{w}_1, \mathbf{w}_2, \beta_{z,k}} \quad & R_{11} + R_{22} \label{eq:opt_problem_a}  \\
\text{s.t.} \quad \quad  
& \| \mathbf{w}_1 \|^2 + \| \mathbf{w}_2 \|^2 \leq P_{\rm max}, \label{eq:opt_problem_b} \\[2pt]
& R_{12} \geq R_{22}, \label{eq:opt_problem_c} \\[2pt]
& \beta_{r,k} + \beta_{t,k} + \beta_{j,k} = 1, 
\quad \forall k \in \mathcal{K}, \label{eq:opt_problem_e} \\[2pt]
& \beta_{z,k} \in \{0,1\}, 
\quad \forall k \in \mathcal{K}, \; z \in \{r,t,j\}, \label{eq:opt_problem_f} \\[2pt]
& \max R_{\rm e1} \leq \tau, \label{eq:opt_problem_g} \\[2pt]
& \theta_{r,k}, \theta_{t,k}, \theta_{j,k} \in [0, 2\pi), 
\quad \forall k \in \mathcal{K}. \label{eq:opt_problem_h}
\end{align}
\end{subequations}

The constraint \eqref{eq:opt_problem_b} enforces the BS transmit power limit, while \eqref{eq:opt_problem_c} ensures the feasibility of NOMA decoding order at User~1 for SIC. 
The constraints \eqref{eq:opt_problem_e}–\eqref{eq:opt_problem_f} define the STAR-RIS element mode selection and exclusivity conditions. 
The constraint \eqref{eq:opt_problem_g} limits the information leakage to Eve, i.e. the maximum allowed value should be less than or equal to $\tau$. 
Finally, \eqref{eq:opt_problem_h} defines the feasible range for the STAR-RIS phase-shift variables.

\vspace{-0.05in}
\section{Proposed Solution}

The optimization problem in (\( \mathrm{P1} \)) is non-convex due to variable coupling, namely the bilinear relation between beamforming vectors and STAR-RIS phase matrices,  and discrete mode-selection coefficients \( \beta_{z,k} \). To address this, we adopt a dual stage penalty optimization that decomposes the problem into two interdependent subproblems, active and passive beamforming, solved iteratively by optimizing one variable set while keeping the other fixed.

\vspace{-0.15in}
\subsection{Active Beamforming Optimization}

In the first stage of the dual-stage framework, we optimize the active beamforming vectors $\mathbf{w}_1 $ and $\mathbf{w}_2 $  at the BS while keeping the STAR-RIS phase-shift matrices \( \boldsymbol{\Theta}_{\mathrm{r}} \), \( \boldsymbol{\Theta}_{\mathrm{t}} \), and \( \boldsymbol{\Theta}_{\mathrm{j}} \) fixed. The objective is to maximize the achievable sum-rate of legitimate users while ensuring the physical layer security constraint against the eavesdropper (Eve) is satisfied.

\textbf{1) Simplification of the objective function by auxiliary variables $r_{11}, r_{22},$ and $s$} : To simplify the formulation, we introduce the lifted matrices \( \mathbf{W}_1 = \mathbf{w}_1 \mathbf{w}_1^H \) and \( \mathbf{W}_2 = \mathbf{w}_2 \mathbf{w}_2^H \), where \( \mathbf{W}_1, \mathbf{W}_2 \in \mathbb{C}^{N \times N} \). These matrices represent the covariance structure of the beamforming vectors, allowing the optimization to be expressed in matrix form. Each \( \mathbf{W}_m \) is Hermitian, positive semidefinite, and rank-one. The active beamforming problem can be formulated as
\begin{subequations} 
\label{eq:ao_subproblem} 
\begin{align} 
\max_{\mathbf{W}_1,\, \mathbf{W}_2} \quad & R_{11} + R_{22} \label{eq:ao_subproblem_obj} \\ 
\text{s.t.} \quad \quad 
&\mathbf{W}_1 \succeq \mathbf{0}, \quad \mathbf{W}_2 \succeq \mathbf{0} \label{eq:9b} \\ 
&\operatorname{rank}(\mathbf{W}_1) = 1 \label{eq:9c} \\ 
&\operatorname{rank}(\mathbf{W}_2) = 1 \label{eq:9d} \\ 
&  \eqref{eq:opt_problem_b}, \eqref{eq:opt_problem_c},\eqref{eq:opt_problem_g}, \eqref{eq:opt_problem_h}. \label{eq:9e}
\end{align} 
\end{subequations}
Note that the constraints in (\ref{eq:9e}) will be considered in terms of the lifted matrices later.

The objective function \( R_{11} + R_{22} \), representing the total achievable sum-rate, is non-convex since each rate term \( R_{mm} = \log_2(1 + \mathrm{SINR}_{mm}) \) involves a non-convex SINR expression with a fractional quadratic dependence on \( \mathbf{W}_1 \) and \( \mathbf{W}_2 \). To reformulate this in a convex form, we introduce slack variables \( r_{11} \) and \( r_{22} \) such that \( R_{mm} = \log_2(1 + r_{mm}) \) and \( r_{mm} \leq \mathrm{SINR}_{mm} \). Observe that the scaled objective function can be written as $\tfrac{1}{2}(R_{11} + R_{22}) \geq \log\!\big(\sqrt{(1 + R_{11})(1 + R_{22})}\big)$. Exploiting the monotonicity of the logarithmic function, the objective can be equivalently expressed in terms of  another auxiliary variable $s$:
\begin{equation}
s \leq \sqrt{(1 + r_{11})(1 + r_{22})}, \label{eq:15}
\end{equation}
  The optimization problem in terms of $s$ will be shown later as (P2) in (\ref{eq:activebeamforming}).
  
Next, the power constraint is expressed in terms of the lifted matrices as
\begin{equation}
\operatorname{Tr}(\mathbf{W}_1) + \operatorname{Tr}(\mathbf{W}_2) \leq P_{\max}. \label{eq:16}
\end{equation}
The rank-one constraints in \eqref{eq:9c} and \eqref{eq:9d} are relaxed resulting in  semidefinite relaxation (SDR).
To simplify channel dependencies, we define equivalent effective channel matrices as:
$\mathbf{H}_1 = \mathrm{diag}(\mathbf{h}_{r1}^H)\mathbf{H}_{br}\mathbf{h}_{b1}^H$, 
$\mathbf{H}_2 = \mathrm{diag}(\mathbf{h}_{t2}^H)\mathbf{H}_{br}$, and 
$\mathbf{H}_e = \mathrm{diag}(\mathbf{h}_{re}^H)\mathbf{H}_{br}\mathbf{h}_{be}^H$. And lifted matrices are: 
 \( \mathbf{R} = \boldsymbol{\phi}_{\mathrm{r}}\boldsymbol{\phi}_{\mathrm{r}}^H \), \( \mathbf{T} = \boldsymbol{\phi}_{\mathrm{t}}\boldsymbol{\phi}_{\mathrm{t}}^H \) and \( \mathbf{J} = \boldsymbol{\phi}_{\mathrm{j}}\boldsymbol{\phi}_{\mathrm{j}}^H \) where $\boldsymbol{\phi}_{\mathrm{r}}$, $\boldsymbol{\phi}_{\mathrm{t}}$, and $\boldsymbol{\phi}_{\mathrm{j}}$ are the vectors containing the diagonal entries of ${\boldsymbol{\Theta}}_{\rm{r}}$, ${\boldsymbol{\Theta}}_{\rm{t}}$, and ${\boldsymbol{\Theta}}_{\rm{j}}$, respectively. 
Using these definitions, the SIC decoding order constraint in \eqref{eq:opt_problem_c} can be expressed as
\begin{equation}
\operatorname{Tr}(\mathbf{H}_1 \mathbf{W}_1 \mathbf{H}_1^H \mathbf{R}) 
\leq \operatorname{Tr}(\mathbf{H}_2 \mathbf{W}_2 \mathbf{H}_2^H \mathbf{T}). 
\label{eq:18}
\end{equation}

\textbf{2) Simplification of the constraints (\ref{eq:opt_problem_c}) and (\ref{eq:opt_problem_g}) by the slack variable $r_{12}$}: The constraint \eqref{eq:opt_problem_c} ensures the SIC decoding order, i.e., User~1 must be able to decode User~2’s message before decoding its own. Mathematically, this translates to the inequality \( R_{12} \geq R_{22} \), or equivalently \( \mathrm{SINR}_{12} \geq \mathrm{SINR}_{22} \). However, the SINR expressions are non-convex and cannot be directly handled by convex solvers. 
To convexify this constraint, we introduce a slack variable \( r_{12} \leq \mathrm{SINR}_{12} \), representing the effective SINR of User~1 when decoding User~2’s signal. Using this variable, the rate-ordering constraint in \eqref{eq:opt_problem_c} can be equivalently expressed as:
\begin{equation}
r_{12} \geq r_{22}, \label{eq:rate}
\end{equation}
where \( r_{22} \) corresponds to User~2’s achievable SINR. This transformation preserves the SIC condition while enabling convex reformulation of the fractional SINR structure.

The instantaneous SINR for User~1 decoding User~2's message is given by
\begin{equation}
\mathrm{SINR}_{12}
= \frac{
    \operatorname{Tr}\!\big( \mathbf{H}_1 \mathbf{W}_2 \mathbf{H}_1^H \mathbf{R} \big)
}{
    \operatorname{Tr}\!\big( \mathbf{H}_1 \mathbf{W}_1 \mathbf{H}_1^H \mathbf{R} \big)
    + \operatorname{Tr}\!\big( \mathbf{H}_1 \mathbf{W}_1 \mathbf{H}_1^H \mathbf{J} \big)
    + \sigma^2
}.
\label{eq:SINR12}
\end{equation}

Rewriting \eqref{eq:SINR12} using \( r_{12} \leq \mathrm{SINR}_{12} \) we get
\begin{equation}
\begin{aligned}
r_{12}
\Big(
    \operatorname{Tr}\!\big( \mathbf{H}_1 \mathbf{W}_1 \mathbf{H}_1^H \mathbf{R} \big)
    + & \operatorname{Tr}\!\big( \mathbf{H}_1 \mathbf{W}_1 \mathbf{H}_1^H \mathbf{J} \big)
    + \sigma^2
\Big)
\le \\
& \operatorname{Tr}\!\big( \mathbf{H}_1 \mathbf{W}_2 \mathbf{H}_1^H \mathbf{R} \big).
\end{aligned}
\label{eq:productform}
\end{equation}

\begin{align}
A &\triangleq \operatorname{Tr}\!\big( \mathbf{H}_1 \mathbf{W}_2 \mathbf{H}_1^H \mathbf{R} \big), \\[4pt]
\Gamma &\triangleq \operatorname{Tr}\!\big( \mathbf{H}_1 \mathbf{W}_1 \mathbf{H}_1^H \mathbf{R} \big)
    + \operatorname{Tr}\!\big( \mathbf{H}_1 \mathbf{W}_1 \mathbf{H}_1^H \mathbf{J} \big)
    + \sigma^2.
\end{align}
Then \eqref{eq:productform} can be rewritten as
\begin{equation}
r_{12}\, \Gamma \le A.
\label{eq:fracineq}
\end{equation}
Applying the arithmetic--geometric mean (AGM) inequality to the product term 
$\tau_{12}\Gamma$, for any auxiliary variable $\mu_{12} = \sqrt{\frac{r_{12}}{\tau}}
$, yields
\begin{equation}
\tau_{12}\Gamma 
\le \frac{1}{2} \left\{ 
(\mu_{12}\Gamma)^2 + \left( \frac{r_{12}}{\mu_{12}} \right)^2\right\}.
\label{eq:agm}
\end{equation}

Combining \eqref{eq:fracineq} and \eqref{eq:agm}, we obtain
\begin{equation}
2A \ge (\mu_{12}\Gamma)^2 + \left( \frac{r_{12}}{\mu_{12}} \right)^2.
\end{equation}

Substituting back $A$ and $\Gamma$, we get
\begin{equation}
\scalebox{0.9}{$
\begin{aligned}
& 2\, \operatorname{Tr}\!\big( \mathbf{H}_1 \mathbf{W}_2 \mathbf{H}_1^H \mathbf{R} \big)
\ge \\
&
\left[
    \mu_{12}
    \Big(
        \operatorname{Tr}\!\big( \mathbf{H}_1 \mathbf{W}_1 \mathbf{H}_1^H \mathbf{R} \big)
        + \operatorname{Tr}\!\big( \mathbf{H}_1 \mathbf{W}_1 \mathbf{H}_1^H \mathbf{J} \big)
        + \sigma^2
    \Big)
\right]^2
+
\left( \frac{r_{12}}{\mu_{12}} \right)^2
\end{aligned}
$}
\label{eq:finalAGM}
\end{equation}

Now, for the term $\left( \frac{r_{12}} {\mu_{12}} \right)^2$,  we add a variable $a$ to ensure numerical stability. To ensure convexity within the second-order cone (SOC) framework, the following SOC constraint is added:
\begin{equation}
\left\| [2r_{12}, \, \mu_{12} - a]^T \right\| \leq \mu_{12} + a. 
\label{eq:23}
\end{equation}
Constraint \eqref{eq:23} equivalently enforces the quadratic inequality \( 4r_{12}a \leq (\mu_{12} + a)^2 - (\mu_{12} - a)^2 \), maintaining convex feasibility. 

Through the combination of the slack variable \( r_{12} \), the monotonic transformation in \eqref{eq:SINR12}, and the AGM–SOC reformulation in \eqref{eq:finalAGM}–\eqref{eq:23}, the original non-convex constraint in \eqref{eq:opt_problem_c} is successfully transformed into a convex representation.
\begin{equation}
2\!\left(\operatorname{Tr}(\mathbf{H}_1 \mathbf{W}_2 \mathbf{H}_1^H \mathbf{R})\right) 
\geq \mu_{12}^2 \Gamma^2 + a,
\label{eq:22}
\end{equation}

The secrecy constraint in \eqref{eq:opt_problem_g} further ensures limited information leakage to Eve, i.e., \( R_{\mathrm{e1}} \leq \tau \), which translates to \( \mathrm{SINR}_{\mathrm{e1}} \leq 2^{\tau} - 1 \). So, $\mathrm{SINR}_{\mathrm{e1}}$ can be now written as 
\begin{equation}
\mathrm{SINR}_{\mathrm{e1}} =
\frac{\operatorname{Tr}\!\left(\mathbf{H}_{\mathrm{e}}\mathbf{W}_1\mathbf{H}_{\mathrm{e}}^H\mathbf{R}\right)}
{\operatorname{Tr}\!\left(\mathbf{H}_{\mathrm{e}}\mathbf{W}_1\mathbf{H}_{\mathrm{e}}^H\mathbf{J}\right) + \sigma^2}.
\end{equation}

The fractional coupling is linearized into the convex constraint
\begin{equation}
\operatorname{Tr}(\mathbf{H}_{\mathrm{e}}\mathbf{W}_1\mathbf{H}_{\mathrm{e}}^H\mathbf{R}) 
\leq \sigma^2 + \operatorname{Tr}(\mathbf{H}_{\mathrm{e}}\mathbf{W}_1\mathbf{H}_{\mathrm{e}}^H\mathbf{J}),
\label{eq:25}
\end{equation}
ensuring that the jamming component dominates at Eve.

Finally, the convexified active beamforming subproblem is formulated as
\begin{align}\label{eq:activebeamforming}
(\mathrm{P2}): \quad 
\max_{\mathbf{W}_1,\, \mathbf{W}_2,\, r_{12},\, s} \quad & s \\[2pt]
\text{s.t.} \quad 
& \eqref{eq:9b},\, \eqref{eq:15},\, \eqref{eq:16},\, \eqref{eq:18}, \nonumber\\
& \eqref{eq:rate},\, \eqref{eq:23},\, \eqref{eq:22},\, \eqref{eq:25}. \nonumber
\end{align}


\vspace{-0.1in}
\subsection{Passive Beamforming Optimization}

In the second stage, the active beamforming matrices \( \mathbf{W}_1 \) and \( \mathbf{W}_2 \) are fixed, and optimization is performed over the STAR-RIS reflection, transmission, and jamming matrices \( \mathbf{R} \), \( \mathbf{T} \), and \( \mathbf{J} \), as well as the mode-selection coefficients \( \beta_{z,k} \). The optimization problem is formulated as
\begin{subequations} \label{eq:26} \begin{align} \max_{s,{\bf{R},J,T},\beta_{z},r_{12}} \quad s \label{eq:ao_subproblem_obj} \\ \text{s.t.} \quad &\mathbf{R} \succeq \mathbf{0}, \quad \mathbf{J} \succeq \mathbf{0}, \quad \mathbf{T}\succeq \mathbf{0} \label{eq:25b} \\ &\operatorname{rank}(\mathbf{R}) = 1, \operatorname{rank}(\mathbf{J}) = 1,\nonumber \\ &\operatorname{rank}(\mathbf{T}) = 1 \label{eq:25c} \\ &\operatorname{Diag}(\mathbf{R}) = \beta_{r}, \operatorname{Diag}(\mathbf{J}) = \beta_{j},\nonumber \\ &\operatorname{Diag}(\mathbf{T}) = \beta_{t} \label{25d} \\ & \eqref{eq:opt_problem_e}, \eqref{eq:opt_problem_f},\eqref{eq:15}, \eqref{eq:18},\eqref{eq:23},\eqref{eq:22},\eqref{eq:25},\end{align}\label{eq:passive} \end{subequations}

This problem remains non-convex primarily due to two sources of nonlinearity: 
(i) the discrete binary nature of the mode-selection variables \( \beta_{z,k} \in \{0,1\} \), and 
(ii) the rank-one constraints imposed on the STAR-RIS matrices \( \mathbf{R}, \mathbf{T}, \mathbf{J} \). 

\textbf{1) Binary mode-selection relaxation:}  
The binary constraint \( \beta_{z,k} \in \{0,1\} \) is equivalently represented by the quadratic inequality 
\( \beta_{z,k}^2 - \beta_{z,k} \leq 0 \) with \( \beta_{z,k} \in [0,1] \). The maximum value of 0 is achieved when $\beta_{z,k}=0$ or $1$. So, we want to maximize \( \beta_{z,k}^2 - \beta_{z,k} \) or minimize \( \beta_{z,k}-\beta_{z,k}^2 \). The non-convex function \( \beta_{z,k}-\beta_{z,k}^2 \) is linearized using the first-order gradient inequality \cite{bertsekas2018nonlinear} at the current iterate \( \beta_{z,k}^{(i)} \):
\begin{equation}
\beta_{z,k} - (\beta_{z,k})^2 
\leq \beta_{z,k} - (\beta_{z,k}^{(i)})^2 - 2\beta_{z,k}^{(i)}(\beta_{z,k} - \beta_{z,k}^{(i)}),
\end{equation}
which provides a tight convex surrogate around the previous iteration.  
To drive each \( \beta_{z,k} \) toward a binary solution, a penalty coefficient \( \zeta > 0 \) is introduced into the objective function. 
The coefficient is progressively increased (\( \zeta \leftarrow \omega \zeta \), \( \omega > 1 \)) after each iteration, ensuring that the solution converges to the discrete set \( \{0,1\} \) as the algorithm proceeds. 

\textbf{2) Rank-one constraint relaxation:}  
The constraints \( \operatorname{rank}(\mathbf{R}) = \operatorname{rank}(\mathbf{T}) = \operatorname{rank}(\mathbf{J}) = 1 \) are non-convex and cannot be handled directly within a semidefinite programming (SDP) framework. 
To approximate the rank function, we employ the difference between the nuclear norm \( \|\mathcal{\bf{Z}}\|_* \) (sum of singular values) and the spectral norm \( \|\mathcal{\mathbf{Z}}\|_2 \) (largest singular value), i.e.,
\begin{equation}
\|\mathcal{\bf{Z}}\|_* - \|\mathcal{\bf{Z}}\|_2 \geq 0, \quad \mathcal{\bf{Z}} \in \{\mathbf{R}, \mathbf{T}, \mathbf{J}\},
\label{eq:nuclear_spectral_diff}
\end{equation}
which is minimized if and only if \( \mathcal{\bf{Z}} \) is rank-one. 
The function in \eqref{eq:nuclear_spectral_diff} is concave with respect to \( \mathcal{\bf{Z}} \) because the spectral norm is convex; therefore, we linearize \( -\|\mathcal{\bf{Z}}\|_2 \) at the current iterate \( \mathcal{\bf{Z}}^{(i)} \) using its subgradient. 
From convex analysis, the subgradient of the spectral norm at \( \mathcal{\bf{Z}}^{(i)} \) is given by the outer product of the dominant left and right singular vectors, denoted by \( \boldsymbol{\lambda}_1(\mathcal{\bf{Z}}^{(i)}) \boldsymbol{\lambda}_1(\mathcal{\bf{Z}}^{(i)})^H \). 
Hence, applying the gradient inequality, we get
\begin{equation}
\scalebox{0.8}{$
\begin{aligned}
\|\mathcal{\bf{Z}}\|_* - \|\mathcal{\bf{Z}}\|_2
\le \operatorname{Tr}(\mathcal{\bf{Z}}) + \|\mathcal{\bf{Z}}^{(i)}\|_2
+ \operatorname{Tr}\!\Big( 
\boldsymbol{\lambda}_1(\mathcal{\bf{Z}}^{(i)})
\boldsymbol{\lambda}_1(\mathcal{\bf{Z}}^{(i)})^H 
(\mathcal{\bf{Z}} - \mathcal{\bf{Z}}^{(i)}) 
\Big)
\end{aligned}
$}
\label{eq:nuclear_norm_ineq}
\end{equation}

This first-order approximation guarantees convexity at each iteration and encourages low-rank (ideally rank-one) solutions for the STAR-RIS matrices. 
A penalty coefficient \( \xi > 0 \) controls the strength of this regularization, ensuring that the obtained solutions asymptotically satisfy the original rank-one constraints.  

By combining the above convex surrogates for the binary and rank-one constraints, the relaxed problem \( \mathrm{(P3)} \) is transformed into a smooth convex program  as shown below:
\begin{equation}
\begin{aligned}
(\mathrm{P3}):~
\max_{s,\mathbf{R},\mathbf{T},\mathbf{J},\boldsymbol{\beta}_z,r_{12}}
\; & s  \\
& - \zeta 
\sum_{z\in\{r,t,j\}}\sum_{k=1}^{K}
\Big[
(\beta_{z,k}-\beta_{z,k}^{(i)})^2
- (\beta_{z,k}^{(i)})^2  \\
& \qquad
-2\beta_{z,k}^{(i)}(\beta_{z,k}-\beta_{z,k}^{(i)})
\Big] \\
& - \xi 
\sum_{\mathcal{\mathbf Z}\in\{\mathbf{R},\mathbf{T},\mathbf{J}\}}
\Big[
\operatorname{Tr}(\mathcal{\mathbf Z})
+ \|\mathcal{\mathbf Z}^{(i)}\|_2  \\
& \qquad
+ \mathcal{Y}(\mathcal{\mathbf Z},\mathcal{\mathbf Z}^{(i)})
\Big] \\
\text{s.t.}\quad
& (13\mathrm{e}),~(15),~(17),~(18),~(29),~(30), \\
& (32),~(34\mathrm{b}),~(34\mathrm{c})
\end{aligned}
\label{eq:passivebeamforming}
\end{equation}
where, $ \mathcal{Y}(\mathbf{Z}, \mathbf{Z}^{(i)}) = \operatorname{Tr}\big(\lambda_1(\mathbf{Z}^{(i)}) \lambda_1(\mathbf{Z}^{(i)})^H (\mathbf{Z} - \mathbf{Z}^{(i)})\big)$

As the penalty coefficients \( \zeta \) and \( \xi \) increase, the solution converges toward the feasible set of the original mixed-integer non-convex problem.
Problems (\( \mathrm{P2} \)) and (\( \mathrm{P3} \)) are solved alternately using the CVX toolbox. After solving (\( \mathrm{P3} \)), a one-hot encoding step is applied where, for each element \( k \), the mode with the highest value among \( \beta_{r,k}, \beta_{t,k}, \beta_{j,k} \) is set to 1 and the others to 0, ensuring that only one mode is active at any time. The penalty coefficients \( \zeta \) and \( \xi \) are gradually increased by a scaling factor \( \omega > 1 \) to enforce convergence toward binary and rank-one solutions. This iterative AO process continues until the improvement in the objective value \( s \) falls below a predefined threshold \( \varepsilon \). 

\vspace{-0.1in}
\begin{algorithm}[!htbp]
\caption{Dual Stage Penalty Optimization }\label{alg:alg1}
\small 
\begin{algorithmic}
\STATE 
\STATE 1: \hspace{0.1cm}$ \textbf{Initialize }  \mathbf{R,T,J},\beta_r,\beta_t,\beta_j,\xi,\zeta,\mu_{12a},\mu_{12b}, \varepsilon,$ set $i=1$
\STATE 2: \hspace{0.5cm}$ \textbf{Repeat}  $
\STATE 4: \hspace{1.5cm} Set $j=1$
\STATE 3: \hspace{1 cm}$ \textbf{Repeat}  $
\STATE 5: \hspace{1.5cm} Solve active beamforming  (\( \mathrm{P2} \)) in (\ref{eq:activebeamforming})
\STATE 6: \hspace{1.5cm}  Update $\mu_{12a}^{(i+1)} = \sqrt{\frac{r_{12}}{\tau}} $ and increment $j$
\STATE 7: \hspace{1 cm} \textbf{Until} $ {|s^{(i+1)} - s^{(i)}|\leq \varepsilon,} $ 
\STATE 8: \hspace{1.5 cm} Output $s^*,{\bf{W}}_1^*,{\bf{W}}_2^*$ \\ \hspace{1.5cm} with optimized values after stage 1
\STATE 9: \hspace{1 cm} Calculate $ R_{1} =  \log_2 \left( (s^*)^2 \right)$
\STATE 10: \hspace{1.5cm} Set  $k=1$
\STATE 11: \hspace{1 cm}$ \textbf{Repeat}  $
\STATE 12: \hspace{1.5cm} Solve passive beamforming  (\( \mathrm{P3} \)) in (\ref{eq:passivebeamforming})
\STATE 13: \hspace{1.5cm}  Update $ \mu_{12b}^{(i+1)} $ and increment $k$
\STATE 14: \hspace{1.5cm}  Update $ \xi = \omega\xi $
\STATE 15: \hspace{1.5cm}  Update $ \zeta = \omega\zeta $
\STATE 16: \hspace{1 cm} \textbf{Until} $ {|s^{(i+1)} - s^{(i)}|\leq \varepsilon,} $
\STATE 17: \hspace{1.5 cm} Output $ s^*,\beta_r^*,\beta_t^*,\beta_j^*,{\bf{R}}^*,{\bf{T}}^*,\mathbf{J}^*, $ \\
\hspace{1.5cm} with optimized values after stage 2. 
\STATE 18: \hspace{1.5cm}  Update $ \beta_r,\beta_t,\beta_j $ by using one hot encoding
\STATE 19: \hspace{1 cm} Calculate $ R_{2} =  \log_2 \left( (s^*)^2 \right)$
\STATE 20: \hspace{0.5 cm} \textbf{Until} $ {|R_{2}- R_{1} |\leq \varepsilon} $

\end{algorithmic}
\label{alg1}
\vspace{-0.05in}
\end{algorithm}

 \vspace{-0.02in}
\section{Results and Discussion}

TABLE~\ref{table:sim_parameters} presents the simulation parameters considered in this paper. Unless stated otherwise, the default system setup assumes a STAR-RIS with \( K = 30 \) elements, a BS equipped with \( N = 2 \) antennas, and perfect CSI at the transmitter is assumed.
Fig.~\ref{fig:fig2} compares the achievable sum-rate of the proposed tertiary-mode STAR-RIS with jamming against several benchmark schemes, including conventional RIS, STAR-RIS without jamming, traditional RIS with jamming, and a baseline without RIS. It is observed that the proposed system consistently outperforms all baselines across the entire transmit power range (0–40 dBm). The STAR-RIS with integrated jamming achieves a notable improvement of approximately 1 bit/s/Hz in secrecy rate over conventional RIS-assisted systems. This performance gain arises from the intelligent mode switching capability of the STAR-RIS, which dynamically allocates elements among transmission, reflection, and jamming functions to maximize secrecy performance. The traditional RIS with jamming and STAR-RIS yield similar results. Their performance is better than that of the traditional RIS. Finally,  systems without RIS exhibit the lowest rates due to their inability to control channel propagation, confirming the significant benefit of incorporating reconfigurable surfaces.

\begin{table}[!htbp]
\centering
\caption{Simulation Parameters}
\label{table:sim_parameters}
\footnotesize
\begin{tabular}{|c|p{3.8cm}|c|}
\hline
\renewcommand{\arraystretch}{1.1}
\textbf{Param.} & \textbf{Description} & \textbf{Value} \\
\hline
$\alpha_{\rm{LOS}} $ & Path loss exponents (Rayleigh) & 2 \\
\hline
$\alpha_{\rm{NLOS}_{\rm r}}$ & Path loss exponents (Rician) & 2.8 \\
\hline
$\alpha_{\rm{NLOS}_{\rm t}}$ & Path loss exponents (Rician) & 3 \\
\hline
$ K_{\mathrm{rician}} $ & Rician factor  & 1 \\
\hline
$\lambda_0$ & Path loss at 1 meter & $-30$ dB \\
\hline
$\varepsilon$ & Convergence threshold  & 0.01 \\
\hline
$\zeta, \xi$ & Penalty factors & 0.0001 \\
\hline
$\omega$ & Scaling factor & 1.5 \\
\hline
$\tau$ & Info leakage tolerance & 1 bit/s/Hz \\
\hline
$\sigma_0$ & Noise power & $-90$ dBm \\
\hline
\end{tabular}
\end{table}
\vspace{-0.1in}

\pgfplotsset{
    every axis/.append style={
        line width=0.9pt,
        tick style={line width=0.9pt},
        width=7.8cm,
        height=6.2cm,
        grid=both,
        grid style={dotted},
        ticklabel style={font=\small,/pgf/number format/fixed},
        label style={font=\small},
        legend style={font=\scriptsize, align=left},
        legend pos=north west
    }
}

\vspace{-0.1in}
\begin{figure}[htb!]
\centering
\begin{tikzpicture}
\begin{axis}[
    xmin=0, xmax=40,
    ymin=0, ymax=15,
    xlabel={\small Transmit Power (dBm)},
    ylabel={\small Rate (bit/s/Hz)},
    ytick={0,2,...,16},
    legend entries={
        STAR-RIS with Jamming,
        STAR-RIS,
        Traditional RIS,
        Traditional RIS with Jamming,
        Without RIS
    }
]
    \addplot[blue,thick,mark=*,mark size=3pt] table [x={x1}, y={y1}] {bnn1.txt};
    \addplot[green,thick,mark=diamond*,mark size=3pt] table [x={x1}, y={y2}] {bnn1.txt};
    \addplot[red,thick,mark=triangle*,mark size=3pt] table [x={x1}, y={y3}] {bnn1.txt};
    \addplot[black,thick,mark=square*,mark size=3pt] table [x={x1}, y={y4}] {bnn1.txt};
    \addplot[cyan,thick,mark=pentagon*,mark size=3pt] table [x={x1}, y={y5}] {bnn1.txt};
\end{axis}
\end{tikzpicture}
\vspace{-0.1cm}
\caption{Sum-rate versus transmit power for different schemes ($K=30$).}
\label{fig:fig2}
\end{figure}
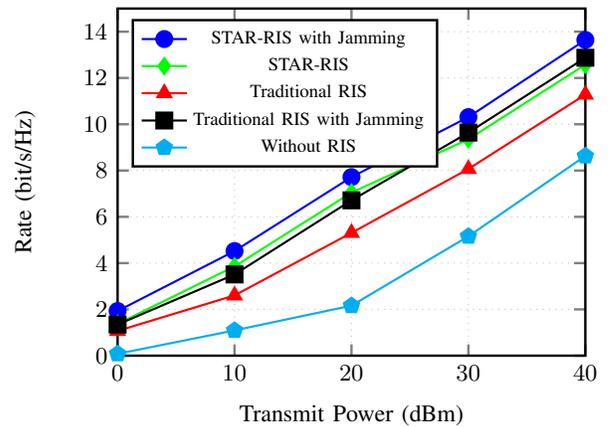
\vspace{-0.1in}
Fig.~\ref{fig:fig3} analyzes the effect of varying the number of STAR-RIS elements (\( K = 20, 30, 40, 50, 60 \)) on the achievable sum-rate. As expected, increasing \( K \) improves the performance since a larger number of elements provides finer control over the incident and reflected wavefronts, leading to higher beamforming gain and better interference management. The improvement becomes less significant beyond \( K = 50 \), suggesting a diminishing return in rate enhancement as the surface size grows. This observation highlights a key trade-off between performance gain and hardware complexity in practical STAR-RIS deployments.
\pgfplotsset{
    every axis/.append style={
        line width=0.9pt,
        tick style={line width=0.9pt},
        width=8cm,
        height=6cm,
        grid=both,
        grid style={dotted},
        ticklabel style={font=\small,/pgf/number format/fixed},
        label style={font=\small},
        legend style={font=\scriptsize, align=left},
        legend pos=north west
    }
}
\vspace{-0.2in}
\begin{figure}[htb!]
\centering
\begin{tikzpicture}
\begin{axis}[
    xmin=0, xmax=40,
    ymin=0, ymax=20,
    xlabel={\small Transmit Power (dBm)},
    ylabel={\small Rate (bit/s/Hz)},
    ytick={0,2,...,20},
    legend style={ legend columns=2},
    legend entries={$K=20$, $K=30$, $K=40$, $K=50$, $K=60$}
]
    \addplot[brown,thick,mark=diamond,mark size=3pt] table [x={x1}, y={y1}] {kvarying.txt};
    \addplot[blue,thick,mark=pentagon,mark size=3pt] table [x={x1}, y={y3}] {kvarying.txt};
    \addplot[red,thick,mark=*,mark size=3pt] table [x={x1}, y={y5}] {kvarying.txt};
    \addplot[green,thick,mark=o,mark size=3pt] table [x={x1}, y={y7}] {kvarying.txt};
    \addplot[black,thick,mark=square,mark size=3pt] table [x={x1}, y={y9}] {kvarying.txt};
\end{axis}
\end{tikzpicture}
\vspace{-0.1cm}
\caption{Impact of STAR-RIS elements $K$ on the sum-rate versus transmit power performance.}
\label{fig:fig3}
\end{figure}
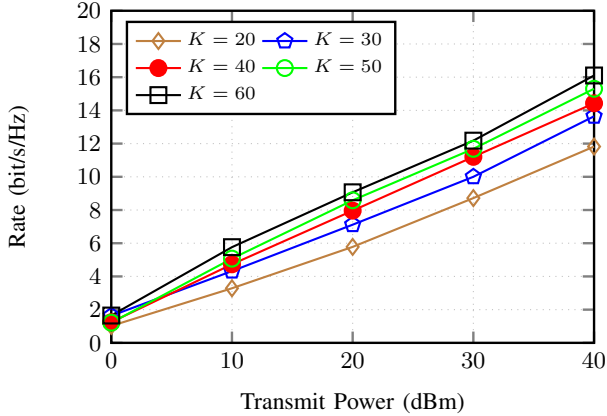
\vspace{-0.1in}

Fig.~\ref{fig:fig5} shows the adaptive mode allocation of STAR-RIS elements (reflection, transmission, and jamming) under different transmit power levels. At lower transmit powers, a majority of the elements operate in the reflection and transmission modes to strengthen the useful signals for legitimate users. However, as the transmit power increases, the system allocates a larger fraction of elements to the jamming mode to effectively suppress potential eavesdropping. The proposed mode-switching framework enables the STAR-RIS to autonomously balance between data forwarding and interference generation.
\pgfplotsset{
    every axis/.append style={
        line width=0.9pt,
        tick style={line width=0.9pt},
        width=8cm,
        height=6cm,
        grid=both,
        grid style={dotted},
        ticklabel style={font=\small,/pgf/number format/fixed},
        label style={font=\small},
        legend style={font=\scriptsize, align=left},
        legend pos=north west
    }
}
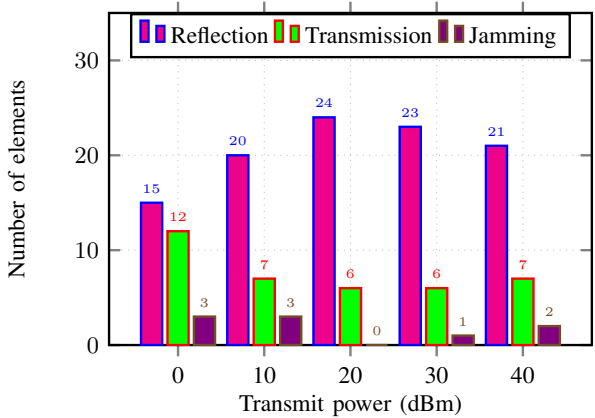
\begin{figure}[htb!]
\centering
\begin{tikzpicture}
\begin{axis}[
    ybar,
    bar width=8pt,
    xlabel={\small Transmit power (dBm)},
    ylabel={\small Number of elements},
    symbolic x coords={0,10,20,30,40},
    xtick=data,
    enlarge x limits=0.2,
    ymin=0, ymax=35,
    grid=both,
    nodes near coords,
    every node near coord/.append style={font=\tiny},
    legend style={at={(0.5,1)}, anchor=north, legend columns=3, font=\small},
    legend entries={Reflection, Transmission, Jamming}
]
    \addplot+[style={fill=magenta}] coordinates {(0,15) (10,20) (20,24) (30,23) (40,21)};
    \addplot+[style={fill=green}] coordinates {(0,12) (10,7) (20,6) (30,6) (40,7)};
    \addplot+[style={fill=violet}] coordinates {(0,3) (10,3) (20,0) (30,1) (40,2)};
\end{axis}
\end{tikzpicture}
\vspace{-0.1cm}
\caption{Distribution of STAR-RIS modes (reflection, transmission, jamming) versus transmit power.}
\label{fig:fig5}
\end{figure}

Fig.~\ref{fig:fig6} demonstrates the convergence performance of the proposed penalty-based alternating optimization algorithm. The figure compares the convergence curves for active beamforming, passive beamforming, and the overall joint optimization process at a transmit power of 40 dBm. It is evident that the algorithm converges within a few iterations (typically less than 10) for all subproblems, indicating computational efficiency. The overall rate improvement stabilizes after approximately six iterations, confirming the stability and robustness of the proposed optimization framework. This demonstrates that the penalty-based method provides fast and stable convergence while maintaining a near-optimal trade-off between rate maximization and secrecy constraints.

\pgfplotsset{
    every axis/.append style={
        line width=0.9pt,
        tick style={line width=0.9pt},
        width=8cm,
        height=6cm,
        grid=both,
        grid style={dotted},
        ticklabel style={font=\small,/pgf/number format/fixed},
        label style={font=\small},
        legend style={font=\scriptsize, align=left},
        legend pos=north west
    }
}
\begin{figure}[htb!]
\centering
\begin{tikzpicture}
\begin{axis}[
    xmin=0, xmax=10,
    ymin=0, ymax=15,
    xlabel={\small Number of Iterations},
    ylabel={\small Rate (bit/s/Hz)},
    xtick={0,1,...,10},
    ytick={0,1,...,15},
    legend style={at={(0.3,0.6)}, legend columns=1},
    legend entries={Convergence for Active Beamforming, Convergence for Passive Beamforming, Convergence for overall algorithm}
]
    \addplot[red,thick,mark=triangle,mark size=3pt] table [x={x1}, y={y1}] {convergence.txt};
    \addplot[green,thick,mark=o,mark size=3pt] table [x={x1}, y={y2}] {convergence.txt};
    \addplot[blue,thick,mark=diamond,mark size=3pt] table [x={x1}, y={y3}] {convergence.txt};
\end{axis}
\end{tikzpicture}
\vspace{-0.3cm}
\caption{Convergence rate of the proposed algorithm for the STAR-RIS system at 40 dBm.}
\label{fig:fig6}
\end{figure}
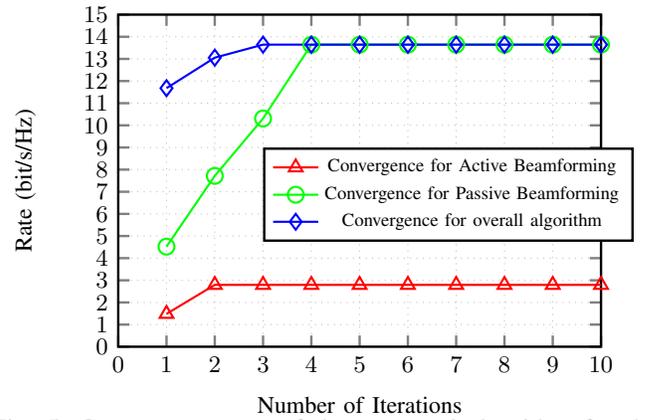

\vspace{-0.2in}

\section{Conclusion}


This paper presented a secure STAR-RIS-aided NOMA system where each STAR-RIS element operates in reflection, transmission, or jamming mode. By jointly optimizing active beamforming at the BS and passive beamforming with mode selection at the STAR-RIS, significant improvements in secrecy and spectral efficiency were achieved under perfect CSI. Simulations showed that combining reflection and jamming provides superior performance compared to conventional IRS-based and non-optimized systems. Future work can extend this framework to scenarios with imperfect CSI, user mobility, or distributed STAR-RIS, and investigate learning-based methods for real-time adaptation in dynamic environments.


\vspace{-0.1in}
\bibliography{Reference}
\bibliographystyle{IEEEtran}

\vfill

\end{document}